\documentclass[11pt,twocolumn]{article} 
\pdfoutput=1

\usepackage[margin={1.5cm,1.5cm}]{geometry} 
\usepackage[all]{nowidow}
\usepackage[version=3]{mhchem}

\usepackage{balance,float}  
\usepackage{subscript,gensymb,amsmath,graphicx}
\usepackage[super,sort&compress,comma]{natbib}

\newcommand{\lo}[1]{\textsubscript{#1}} 
\newcommand{\hi}[1]{\textsuperscript{#1}}

\begin{document}

\title{Combinatorial Insights into Doping Control and Transport Properties of Zinc Tin Nitride}

\author{Angela N. Fioretti,$^{1,2~\ast}$ Andriy Zakutayev,$^{1}$ Helio Moutinho,$^{1}$ Celeste Melamed,$^{3}$\\ 
John D. Perkins,$^{1}$ Andrew G. Norman,$^{1}$ Mowafak Al-Jassim,$^{1}$\\ 
Eric S. Toberer,$^{1,2}$ Adele C. Tamboli$^{1,2}$\\
\\
\normalsize{$^{1}$National Renewable Energy Laboratory, Golden, Colorado 80401 USA}\\
\normalsize{$^{2}$Colorado School of Mines, Golden, Colorado 80401 USA}\\
\normalsize{$^{3}$Harvey Mudd College, Claremont, California 91711 USA}\\
\\
\normalsize{$^\ast$To whom correspondence should be addressed; E-mail: afiorett@mines.edu.}}

\date{} 

\twocolumn[
\begin{@twocolumnfalse}
\maketitle
\renewcommand{\abstractname}{Broader Context}
\begin{abstract}
\noindent{Thin film photovoltaics (PV), based on materials that absorb light 10--100 times more efficiently than crystalline silicon, has the potential to drive down PV module cost by increasing efficiency and requiring less raw material. Currently, the market for thin film PV is dominated by CdTe and CuIn\lo{x}Ga\lo{1-x}Se\lo{2} technologies, which suffer from concerns of toxicity (Cd) or rarity (Te, In) when considering terawatt-scale deployment. As an alternative to these technologies, researchers have turned to studying new absorber materials that exhibit equivalent light absorbing properties but are comprised of Earth-abundant elements. In this work, we explore the properties of a thin film III-N analog, ZnSnN\lo{2}, that until recently has received little attention in the literature. Classification as a III-N analog is advantageous for PV applications, considering that III-N materials are well-known for their stability. ZnSnN\lo{2} possesses a direct bandgap and large absorption coefficient in a PV-relevant energy range, in addition to being composed of abundant and non-toxic elements. However, a few key optoelectronic properties (\emph{e.g.} doping control and exact bandgap) of this material are not well understood. If this challenge is addressed, ZnSnN\lo{2} has excellent potential as an absorber material for Earth-abundant thin film PV.} 
\end{abstract}
\renewcommand{\abstractname}{Abstract}
\begin{abstract}
\noindent{ZnSnN\lo{2} is an Earth-abundant semiconductor analogous to the III-Nitrides with potential as a solar absorber due to its direct bandgap, steep absorption onset, and disorder-driven bandgap tunability. Despite these desirable properties, discrepancies in the fundamental bandgap and degenerate \emph{n}-type carrier density have been prevalent issues in the limited amount of literature available on this material. Using a combinatorial RF co-sputtering approach, we have been able to explore a growth-temperature-composition space for Zn\lo{1+x}Sn\lo{1-x}N\lo{2} over the ranges 35--340\degree C and 0.30--0.75 Zn/(Zn+Sn). In this way, we were able to identify an optimal set of deposition parameters for obtaining as-deposited films with wurtzite crystal structure and carrier density as low as 1.8 x 10\hi{18} cm\hi{-3}. Films grown at 230\degree C with Zn/(Zn+Sn) = 0.60 were found to have the largest grain size overall (70 nm diameter on average) while also exhibiting low carrier density (3 x 10\hi{18} cm\hi{-3}) and high mobility (8.3 cm\hi{2} V\hi{-1} s\hi{-1}). Furthermore, we report tunable carrier density as a function of cation composition, in which lower carrier density is observed for higher Zn content. This relationship manifests as a Burstein-Moss shift widening the apparent bandgap as cation composition becomes increasingly Zn-poor. Collectively, these findings provide important insight into the fundamental properties of the Zn-Sn-N material system and highlight the potential to utilize ZnSnN\lo{2} for photovoltaics.} 
\end{abstract}
\vspace{0.5cm}
\end{@twocolumnfalse}
]

\section{Introduction}

ZnSnN\lo{2} is an Earth-abundant III-N analog with large optical absorption coefficient and a direct bandgap reported in the highly desirable 1.0--2.0 eV range.\cite{chen2014, punya2011pssc, punya2011prb, quayle2013, feldberg2013} The bandgap of ZnSnN\lo{2} is predicted to be tunable over that range based on degree of cation disorder, making this material potentially useful for many optoelectronic applications.\cite{feldberg2013}  Recent predictions\cite{lahourcade2013} have indicated that the fundamental gap of ZnSnN\lo{2} may be 1.4 eV; remarkably close to the optimal single junction value in the detailed balance limit.\cite{queisser1961, tiedje1984} ZnSnN\lo{2} also holds significant promise for tandem solar cell applications. The predicted bandgap tunability for this material includes the 1.6--1.7~eV bandgap range identified for the top cell in a two-junction device.\cite{bremner2008} In addition to these compelling characteristics, ZnSnN\lo{2} is also part of a trio of materials (Zn-IV-N\lo{2}; IV = Si, Ge, Sn) that possess bandgaps spanning the visible spectrum (from 1.0--5.0 eV).\cite{punya2011pssc, punya2011prb, quayle2013} Alloys of this system are considered to be Earth-abundant alternatives to the InGaN system, due to their lower formation enthalpy (thus potentially avoiding phase segregation) and their solar-relevant bandgap range.\cite{narang2014,coronel2012} It is this suite of properties that has piqued recent interest in ZnSnN\lo{2} and the related Zn-IV-N\lo{2} family for photovoltaics (PV). 

Despite the many potential benefits of this material, ZnSnN\lo{2} is one of the least-studied members of the II-IV-V\lo{2} class.\cite{pamplin1974,goodman1991,shaposhnikov2012} Computational research into its properties did not begin until 2008,\cite{paudel2008} and the first synthesis of ZnSnN\lo{2} was not reported until 2013.\cite{lahourcade2013} Since then, synthesis of ZnSnN\lo{2} has been reproduced, but questions remain concerning both its structure and its fundamental properties.\cite{lahourcade2013, feldberg2013, quayle2013, feldberg2014, deng2015} 

One critical challenge frustrating initial ZnSnN\lo{2} development for optoelectronics was degenerate \emph{n}-type carrier density consistently found in stoichiometric samples. Carrier densities were typically greater than 10\hi{20} cm\hi{-3}, likely due to poor nitrogen incorporation or the presence of oxygen impurities that are difficult to avoid in nitride materials.\cite{lahourcade2013, feldberg2013, quayle2013, feldberg2014} However, with the publication of Ref. [17]\nocite{deng2015} and the results of the present study, it is now known that carrier densities on the order of 10\hi{17}--10\hi{18} cm\hi{-3} are attainable, which approaches a relevant range for PV applications. 

Still, major discrepancies in the value of the fundamental bandgap and crystal structure are common in the ZnSnN\lo{2} literature. Calculated bandgap values range from 0.35--2.64 eV depending on the approximation used and depending on the crystal structure assumed (\emph{i.e.} ordered orthorhombic vs. cation-disordered wurtzite).\cite{punya2011pssc, punya2011prb, feldberg2013} Experimental results are limited, with measured bandgaps ranging from 1.7--2.1 eV.\cite{lahourcade2013,quayle2013,deng2015} These discrepancies between theory and experiment have been attributed to band-filling due to degenerate carrier density\cite{lahourcade2013,feldberg2013} or to changes in cation ordering altering the fundamental gap.\cite{feldberg2013} The degree of cation ordering has been linked to bandgap tuning in related II-IV-V\lo{2} materials, such as ZnSnP\lo{2}, and is expected to affect the ZnSnN\lo{2} bandgap as well.\cite{scanlon2012} However, it has been difficult to determine which hypothesis is correct, or if it is a combination of both effects, because the body of work available on ZnSnN\lo{2} remains quite limited. 

In this work, we used a combinatorial approach to identify a practical range of growth temperatures and cation compositions for growing cation-disordered ZnSnN\lo{2}. We find [0001] growth in the wurtzite structure and present the first report of doping control in ZnSnN\lo{2} through varying cation off-stoichiometry. We find that a 20\% zinc-rich cation composition yields an \emph{n}-type carrier density of 1.8~x~10\hi{18} cm\hi{-3}; the lowest carrier density achieved thus far for as-deposited films. This finding suggests defect compensation or complexing must be occurring in this material. Additionally, we report ZnSnN\lo{2} films that luminesce at cryogenic temperatures, and exhibit no detectable free carrier absorption below the absorption edge at room temperature. These findings not only provide important insight into the fundamental properties of the Zn-Sn-N material system, but also highlight the potential to utilize ZnSnN\lo{2} as a PV absorber material.

\section{Experimental}

\textbf{Combinatorial Experiments.} A combinatorial approach\cite{green2013} was used to deposit thin film libraries of ZnSnN\lo{2} with orthogonal gradients in cation composition and growth temperature. Films were 200--700 nm thick and were deposited from 50 mm diameter elemental Zn and Sn targets of 99.995\% and 99.998\% purity, respectively, via reactive radio frequency (RF) sputtering. Target power was kept at 35 W for Zn and 25 W for Sn, to compensate for the higher volatility of Zn. An RF-plasma atomic nitrogen source (HD25, Oxford Applied Research) run at 250~W and flowing 10 sccm of 99.999\% purity N\lo{2} gas was used to provide the reactive N* species for all films grown in this work.\cite{caskey2014} Films were deposited on 50~x~50 mm Eagle XG glass substrates located 13 cm from the targets in a chamber with 10\hi{-6} Torr residual water base pressure. Depositions were performed in a mixed Ar and N* atmosphere with 10 mTorr of each for a nominal chamber pressure of 20 mTorr. Two sputter guns were inclined at 45\degree~to the substrate normal to create the continuous composition spread of Zn:Sn for each library. A growth temperature gradient was induced perpendicular to the cation composition gradient by contacting the substrate on one end with a heated metal pad.\cite{subramaniyan2014} Hot-side set point temperatures were 110\degree C, 220\degree C, 330\degree C, and 400\degree C, which provided an overall spread in growth temperature from 340\degree C to 60\degree C. Actual growth temperatures achieved were calibrated by placing a thermocouple in contact with the substrate surface at four representative regions (along the decreasing temperature gradient and spaced 12.5 mm apart) for each of the hot-side temperatures mentioned. Additionally, a sample was grown with no active heating of the substrate, and was calibrated by thermocouple to be at 35\degree C as a result of sputtered particle impingement on the substrate surface during growth. Another isothermal sample was grown at 300\degree C with the same composition gradient as the 330\degree C hot-side sample for non-combinatorial advanced characterization. Prior to each deposition, targets were pre-sputtered for at least 30 min with a shutter covering the substrate. 

Sample libraries were characterized using a 4~x~11 mapping grid to connect observed trends in estimated bandgap, crystal structure, and other properties to changes in film composition or growth temperature as a function of position on the substrate.\cite{zakutayev2013b} Mapping-style characterization was used to collect X-ray diffraction (XRD), X-ray fluorescence (XRF), four-point probe (4pp), and UV-Vis-NIR spectroscopy data. Finally, custom software packages written in Igor Pro were used to parse and analyze the substantial amount of data produced by the combinatorial experiments.

XRD was performed using a $\theta$ -- 2$\theta$ geometry with Cu K-$\alpha$ radiation and a proportional 2D detector on a Bruker D8 Discover equipped with General Area Detector Diffraction System software. XRF was performed on a Fischer XDV-SDD instrument to determine both the Zn/Sn ratios and the thickness of the films. Film thickness from XRF was calibrated with scanning electron microscopy (SEM) measurements (described below). Sheet resistance measurements were performed using a collinear 4pp mapping instrument built in-house with 1~mm spacing between probes. Electrical conductivity was subsequently calculated using thickness data from the XRF measurements. Transmission (\emph{T}) and reflection (\emph{R}) spectra were collected in the UV-Vis-NIR spectral ranges (300--2000 nm) using a home-built thin film optical spectroscopy system equipped with deuterium and tungsten/halogen light sources and Si and InGaAs detector arrays. The collected spectra were then used to calculate absorption coefficient ($\alpha$) using the relation $\alpha = -ln[T/(1-R)]/d$, where \emph{d} is the measured film thickness.\cite{zakutayev2012}

\begin{figure*}[t!]
	\centering
		\includegraphics[width=19cm]{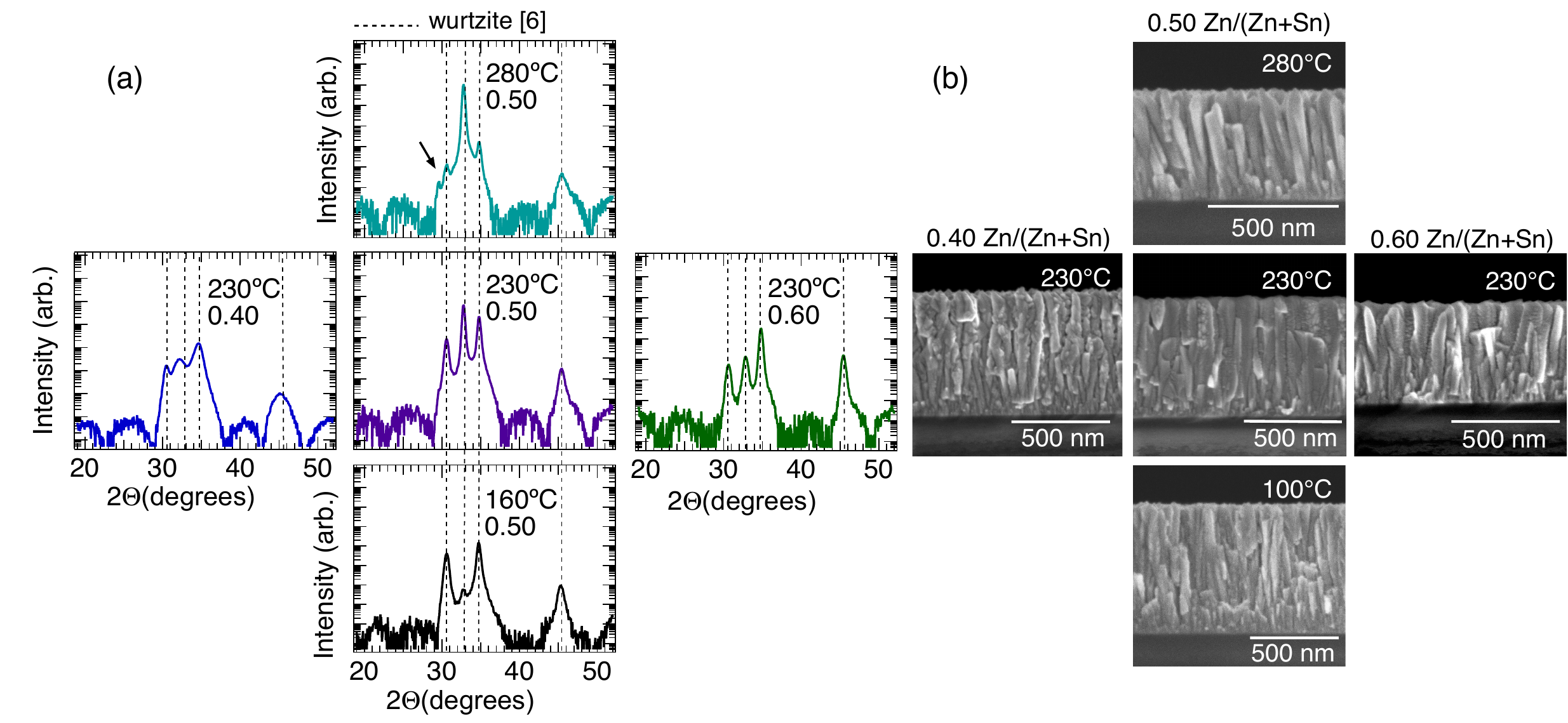}
	\caption{(a) Characteristic XRD patterns for films grown within a broad range of deposition conditions. Numbers in each panel refer to Zn/(Zn+Sn) composition. The black arrow in the top panel of (a) indicates two peaks that are consistent with the Pna2\lo{1} orthorhombic crystal structure.\cite{lahourcade2013} Corresponding cross-sectional SEM images in (b) show columnar grain structure for stoichiometric and zinc-rich films, and more disordered grain structure for tin-rich films.}
\end{figure*}

\textbf{Single Point Experiments.} To obtain morphology, carrier density, and mobility as a function of cation composition and growth temperature, carefully selected regions of each library were manually scribed and cleaved into 7~x~7 mm sections for scanning electron microscopy (SEM) and room temperature Hall effect measurements. These sections were considered to be uniform with regards to composition and morphology due to the small size of the film sections. SEM was performed on an FEI Nova NanoSEM 630 operated at 3 kV and 40 pA to minimize charging effects. Room temperature Hall effect measurements were performed using soldered In contacts in the Van der Pauw configuration on a BioRad HL5500 PC instrument.

Additional single point characterization was performed on specific regions of interest on some combinatorial libraries using the same 7~x~7 mm sample size described above. Transmission electron microscopy (TEM) was performed on thin sections of the 7~x~7 mm film pieces prepared in cross-section in an FEI focussed ion beam (FIB) work station, using the lift-out technique described elsewhere.\cite{norman2012}  Bright-field diffraction contrast TEM and transmission electron diffraction (TED) were  performed at 300 kV in an FEI Tecnai G2 30 S-TWIN TEM. Interplanar spacings were calculated from the TED patterns using a pattern taken from single crystal GaAs for calibrating the spacings given by the instrument under the same measurement parameters. High angle annular dark field (HAADF) imaging paired with energy dispersive X-ray spectroscopy (EDX) analysis was performed in an FEI Tecnai F20 UltraTwin scanning transmission electron microscope (STEM) operated at 200 kV and was used to
analyze the composition of the grain boundaries. Electron backscatter diffraction (EBSD) was performed using an EDAX Hikari A40 system integrated with the same SEM instrument described above. Kikuchi patterns were indexed using both hexagonal and cubic lattice data files to check that the best fit (highest confidence index) happened for the hexagonal lattice. Finally, photoluminescence spectroscopy (PL) was performed at 4.25 K on samples of the 7~x~7 mm size using a 514 nm Modu-Laser Stellar-Pro Argon/Ion laser powered at 10 mW with a 280 $\mu$m slit width and a 550 nm long-pass filter.

To determine oxygen impurity concentration, a nominally stoichiometric sample was grown isothermally at 230\degree C to a thickness of 140 nm for Rutherford backscatter spectrometry (RBS) analysis. This sample was grown on (100) Si to avoid contribution to the oxygen signal from the oxygen in the glass. RBS was performed on a model 3S-MR10 RBS system from National Electrostatics Corporation. The beam consisted of 2 MeV alpha particles at a current of 50 nA. The total accumulated charge was 500 $\mu$C, which indicates a signal integration time of more than 12 times the standard integration time (standard accumulated charge is 40 $\mu$C for this system). The RBS detector was mounted in a 168\degree~backscatter configuration. Analysis of the RBS spectra was performed using RUMP data analysis software,\cite{barradas2008} in which the ZnSnN\lo{2} film was simulated as a layer on top of a 5000 nm thick Si layer for the substrate. To perform the fitting and integration procedures, the oxygen region was excluded when fitting the background signal from the Si substrate, in order to obtain a reliable baseline over which the oxygen signal could later be integrated. Finally, an empirical density of 0.719 x 10\hi{23} atoms cm\hi{-3} was used to calculate thickness (value taken from the NREL Materials Database, entry number 11591).

\section{Results and Discussion}
\subsection{Combinatorial Survey}
\subsubsection{Morphology and Structure}

Using a combinatorial approach, we were able to identify optimal conditions for depositing wurtzite ZnSnN\lo{2} on glass with long range phase-purity by XRD. Six combinatorial libraries were prepared spanning growth temperatures from 35--340\degree C and cation compositions of 0.30--0.75 Zn/(Zn+Sn) as determined by XRF. Throughout this work, the fraction of zinc atomic percent (at\%) to total cation at\% as measured by XRF gives the degree of off-stoichiometry in reference to zinc content. A value of 0.50 Zn/(Zn+Sn) indicates stoichiometric, while a value of 0.60 Zn/(Zn+Sn) indicates zinc-rich. Films falling in the range of 0.45--0.70 Zn/(Zn+Sn) and grown at 120--340\degree C exhibited wurtzite XRD with no peak shift or broadening that would suggest the presence of secondary phases (such as Zn\lo{3}N\lo{2} or Sn\lo{3}N\lo{4}).   

Fig. 1a displays XRD patterns on a log intensity scale for ZnSnN\lo{2} films grown at a range of conditions. At growth temperatures between 160--340\degree C, films with cation compositions of 0.45--0.70 Zn exhibited crystal structure consistent with the ``average wurtzite'' structure observed previously (\emph{i.e.} random cation site-occupancy).\cite{lahourcade2013} The dashed lines in Fig. 1a give the peak positions calculated for wurtzite ZnSnN\lo{2}.\cite{lahourcade2013} Films grown outside this temperature-composition window exhibited either broad XRD peaks that were not well resolved or peaks that were shifted to higher or lower Bragg angle, making it difficult to rule out the presence of secondary phases. These effects can be seen in the far left panel of Fig. 1a (dark blue curve). A set of samples was also prepared isothermally at 400\degree C, but these films showed regions of no net deposition and of metallic zinc and tin. Due to this, 340\degree C was treated as the upper limit on optimal growth temperature.

In contrast to the wurtzite crystal structure observed most commonly in this work, stoichiometric films grown between 280--340\degree C showed evidence of Pna2\lo{1} orthorhombic structure (top panel of Fig. 1a, turquoise curve). Two peaks appear at $\sim$30\degree~$2\theta$, which are at the expected peak positions of the  calculated for ZnSnN\lo{2} in prior work.\cite{lahourcade2013} Off-stoichiometric films grown between 280--340\degree C exhibited wurtzite ZnSnN\lo{2} structure (not shown). Previous calculations have shown that cation-ordered ZnSnN\lo{2} (\emph{i.e.} with orthorhombic crystal structure) should have a bandgap of 2.0 eV, while cation-disordered ZnSnN\lo{2} (with wurtzite crystal structure) should have a bandgap closer to 1.0 eV.\cite{feldberg2013,chen2014} Finding evidence in this work of increased order (orthorhombic crystal structure) at high growth temperature and wurtzite ZnSnN\lo{2} at lower temperature demonstrates the possibility for tuning order parameter, and subsequently bandgap,\cite{feldberg2013,chen2014} through varying growth temperature in this material. 

Indeed, there is a precedent for tuning bandgap and other properties by controlling cation disorder in multinary, tetrahedrally-bonded materials. In one study on ZnSnP\lo{2} (another II-IV-V\lo{2} material), bandgap tuning over 300 meV was achieved through controlling cation disorder by minimally varying the cation flux ratio during growth.\cite{stjean2010} In addition to this study, a theoretical work on ZnSnP\lo{2} showed that bandgap tuning could be achieved over an even wider energy range by varying growth or annealing temperature to control disorder.\cite{scanlon2012} Beyond II-IV-V\lo{2} materials, control of cation disorder to tune properties has been shown in both Cu\lo{2}SnS\lo{3}\cite{baranowski2015} and Cu\lo{2}ZnSnSe\lo{4}.\cite{rey2014} These materials are both tetrahedrally-bonded multinary semiconductors with underlying zincblende lattice that is similar to the underlying wurtzite ZnSnN\lo{2} lattice. For Cu\lo{2}SnS\lo{3}, cation disorder can be controlled through post-growth annealing to reduce point defect density and carrier concentration.\cite{baranowski2015} For Cu\lo{2}ZnSnSe\lo{4}, similar control of cation disorder through anneal/quench cycles was found to reversibly tune the bandgap over 120 meV.\cite{rey2014} Given these examples of controlling cation disorder to tune properties in other multinary, tetrahedrally-bonded materials, it is intriguing that we observe a relationship between degree of disorder and growth temperature in this work, as is observed in the middle column of panels in Fig. 1a. Such a property-disorder relationship, as predicted previously for ZnSnN\lo{2} and shown here to be feasible, could be manipulated to tune the bandgap in this material.\cite{feldberg2013,chen2014}

Films had dense, columnar growth by SEM and exhibited grains with diameter 30--70 nm if grown between 160--340\degree C and 0.45--0.70 Zn/(Zn+Sn) (Fig. 1b, middle, top, and right panels). Deposition rates of up to 300 nm/hr were achieved at all deposition conditions up to 340\degree C. The largest grains observed were 70 nm in diameter on average, for films grown at 230\degree C with 0.60 Zn on the cation site (far right panel of Fig. 1b). Films with far Sn-rich composition (\emph{i.e.} \textless 0.45 Zn on the cation site) showed disrupted columnar growth and rough grain boundaries by SEM (far left panel of Fig. 1b). Columnar growth was also disrupted for films with \textgreater 0.70 Zn on the cation site grown in the 160--340\degree C range. Zn-rich or stoichiometric films grown at temperatures lower than 160\degree C had small grains with diameters \textless 30 nm and sometimes even \textless 10 nm. Grain size was consistently $\leq$ 70 nm for all films grown in this work, but this could be increased by post-growth annealing or growing on oriented substrates.

We observe surprisingly consistent phase stability for films grown over a wide range of deposition conditions. This is in contrast to the very narrow window of equilibrium phase stability predicted by Ref. [1],\nocite{chen2014} in which Zn\lo{3}N\lo{2} and metallic Zn and Sn secondary phases were predicted to be difficult to avoid. We see no evidence of these secondary phases by XRD within the range of 0.45--0.70 Zn/(Zn+Sn) and 160--340\degree C growth temperature. The absence of secondary phases in our films over such variable growth conditions may be due to growing wurtzite ZnSnN\lo{2} instead of the orthorhombic phase assumed in constructing the equilibrium phase stability diagram.\cite{chen2014}

Synthesis of phase-pure, wurtzite ZnSnN\lo{2} suggests that our growth conditions are not in line with the assumptions used to construct the phase stability diagram cited above. For example, the use of activated nitrogen in the present study provides non-equilibrium growth conditions that have been previously shown to widen the range for stable phase formation in other nitride materials.\cite{caskey2014,caskey2015} For example, prior work on the metastable material Cu\lo{3}N using an atomic nitrogen source showed a distinct broadening of the window where this material could be grown.\cite{caskey2014} For ZnSnN\lo{2}, the defect formation enthalpies reported in Ref. [1]\nocite{chen2014} predict that O\lo{N} and V\lo{N} donor defects are two of the four most favorable defects to form. This indicates that nitrogen-rich growth should lead to better material in terms of point defect density. The ability in this work to grow wurtzite ZnSnN\lo{2} at a wide range of conditions highlights the advantage provided by the use of an activated nitrogen source.

\subsubsection{Electrical Properties}

Conductivity was found to vary over four orders of magnitude as a function of cation composition and growth temperature (Fig. 2a). The dashed vertical line in Fig. 2 indicates nominally stoichiometric cation composition, and the grey shaded regions delimit the cation composition range for which wurtzite XRD with no peak shift or broadening was observed (discussed in section 3.1.1). As Zn content increased from 0.35 to 0.65 Zn/(Zn+Sn), conductivity decreased by a factor of 10\hi{4} to a minimum at 0.1 S cm\hi{-1}. Beyond $\sim$0.65 Zn/(Zn+Sn), conductivity increased again, although at compositions past 0.70 Zn/(Zn+Sn), phase-purity was unclear. We note that Zn content higher than $\sim$0.60 was difficult to achieve for growth above 230\degree C, due to the higher volatility of Zn compared to Sn. For fixed film composition at 0.60 Zn/(Zn+Sn), increasing temperature leads to a decreasing conductivity until 120\degree C (dark blue diamond); beyond this critical temperature, further raising the growth temperature increased the conductivity. For more detail on the effect of growth temperature on the conductivity of Zn-rich films, shown in Fig. S1 of ESI.

\begin{figure}[ht!]
	\centering
		\includegraphics[width=6.5cm]{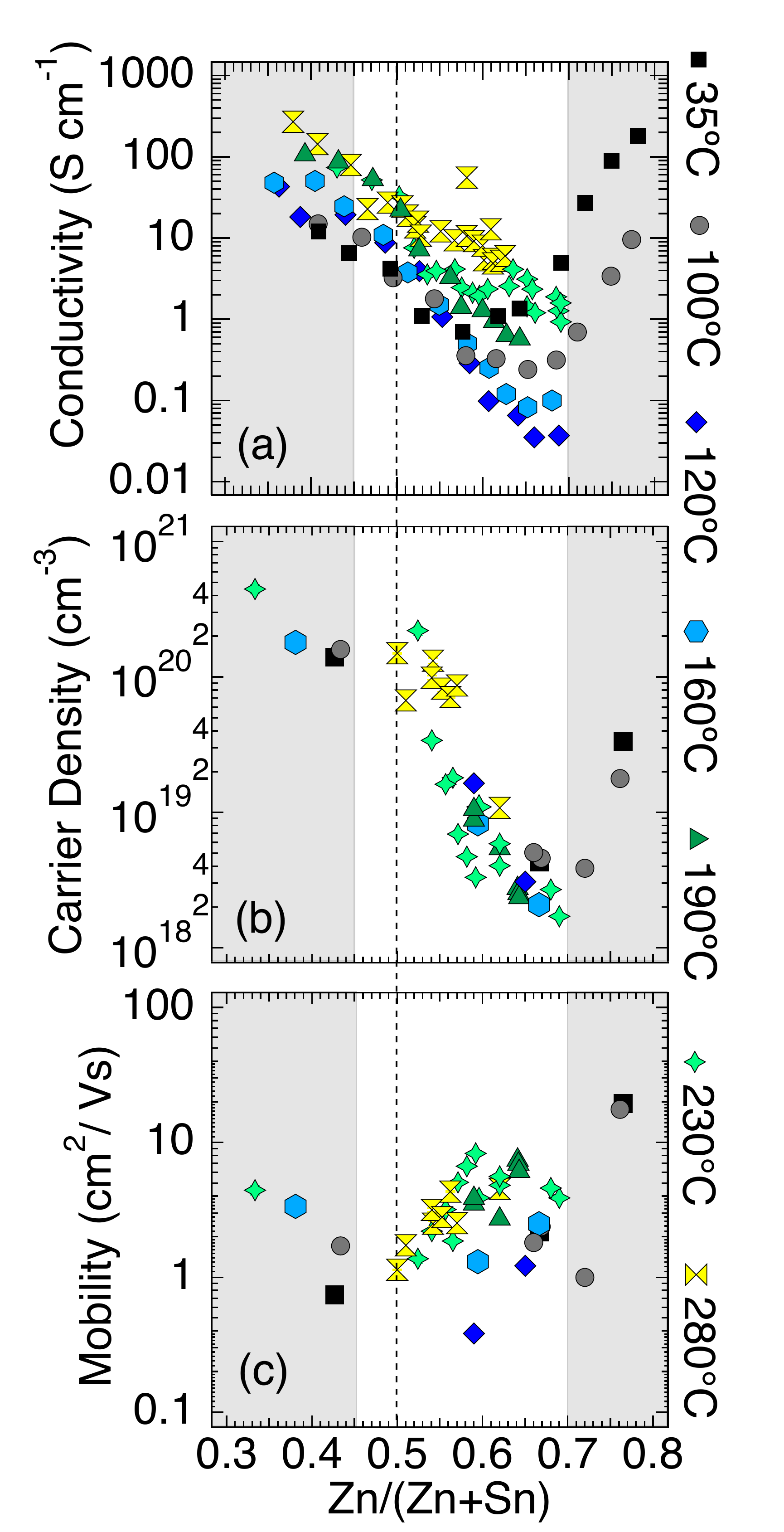}
	\caption{Conductivity measured by 4pp (a) decreased by four orders of magnitude as a function of cation composition. Free electron density measured by room temperature Hall effect (b) decreased by more than two orders of magnitude over the same cation composition range. No trend in this broad data set was observed in mobility as a function of cation composition or growth temperature. The dashed vertical line in Fig. 2 indicates nominally stoichiometric cation composition, and the grey shaded regions delimit the cation composition range for which wurtzite XRD with no peak shift or broadening was observed (discussed in section 3.1.1).}
\end{figure}

Free electron density (Fig. 2b) was found to decrease by more than two orders of magnitude over the same range of Zn content in which conductivity decreased. The lowest \emph{n}-type carrier density observed was 1.8~x~10\hi{18} cm\hi{-3} for a 230\degree C film with 0.70 Zn/(Zn+Sn). For films with the largest grain size (70 nm, 230\degree C, 0.60 Zn/(Zn+Sn)), the lowest carrier density was 3~x~10\hi{18} cm\hi{-3}. Within the optimal range of growth conditions identified, the highest mobility measured was 8.3~cm\hi{2}V\hi{-1}s\hi{-1} and the lowest was 1.1 cm\hi{2}V\hi{-1}s\hi{-1}, although no clear trend in mobility was observed as cation composition varied (Fig. 2c). The mobility values found in this work are on par with, or exceed, what has been previously reported for ZnSnN\lo{2} in literature.\cite{feldberg2013,lahourcade2013,deng2015}

The carrier densities reported here for as-deposited films (low 10\hi{18} cm\hi{-3}) are two orders of magnitude lower than previously reported for ZnSnN\lo{2} films grown without post-growth annealing.\cite{feldberg2013,lahourcade2013} The films in this work exhibit carrier density values comparable to those reported for annealed ZnSnN\lo{2}, in which carrier densities of 10\hi{17}--10\hi{18} cm\hi{-3} were observed.\cite{deng2015} This comparison is promising, because it shows that there are at least two routes to lowering the carrier density in ZnSnN\lo{2}: post-growth annealing or off-stoichiometry. Together, these findings reveal that one of the fundamental challenges for ZnSnN\lo{2}-based photovoltaics, namely degenerate carrier density, may in fact be a tractable problem.	

It is interesting to note that our lowest carrier densities were observed for film compositions around 0.65 Zn/(Zn+Sn), rather than for the stoichiometric 0.50 value. Stoichiometric films, in this work and others,\cite{lahourcade2013, feldberg2013, feldberg2014, quayle2013} exhibit degenerate \emph{n}-type carrier density $\geq$10\hi{20} cm\hi{-3}. The reason for this degeneracy is not obvious, although for stoichiometric cation ratio, it is apparent that degenerate carrier density cannot be easily explained by cation-site defects. Neither can it be explained by Zn\lo{i} defects, which have high formation energy.\cite{chen2014} This suggests that \emph{n}-type degeneracy is due to the anion sublattice, which is consistent with O\lo{N} and V\lo{N} being low energy defects in ZnSnN\lo{2}.\cite{chen2014} Moreover, O\lo{N} and V\lo{N} are well-known point defects in nitride materials in general. The observation that 0.65 Zn/(Zn+Sn) results in non-degenerate films is consistent with this picture, as excess Zn on the disordered cation sublattice would compensate these anion sources of degenerate carrier density. While prior literature identifies Sn\lo{Zn} antisites as the lowest energy defect, that calculation is for a particular chemical potential growth environment. The chemical potential growth environment is clearly varying across the combinatorial libraries grown in this work, making it debatable whether the defect formation energies for orthorhombic ZnSnN\lo{2} can be applied to the wurtzite version found here. This discrepancy highlights the need for future computational efforts on the defect physics of the full Zn-Sn-N material system.

\begin{figure}[h!] 
	\centering
		\includegraphics[width=10cm]{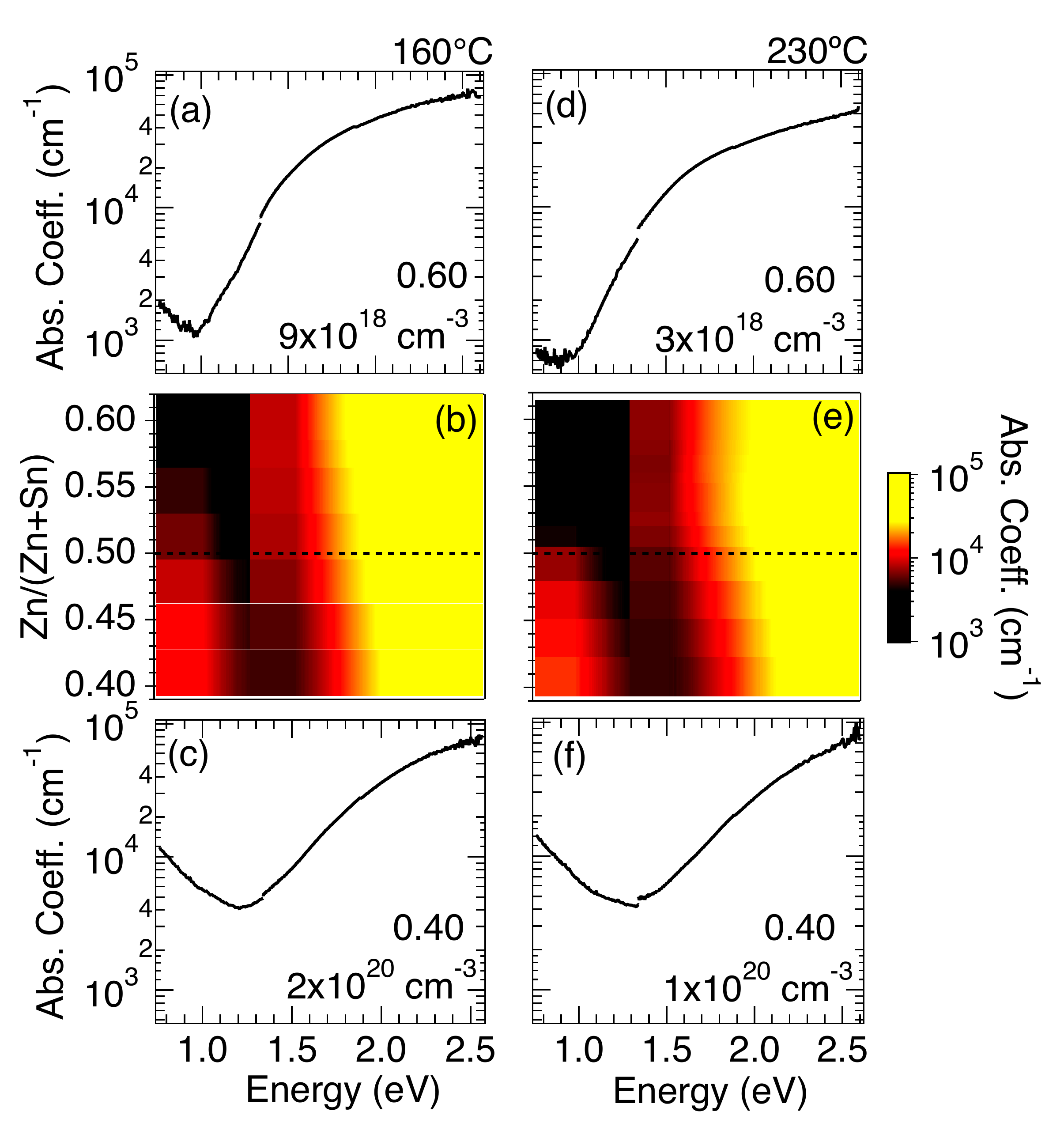}
	\caption{Absorption coefficient plotted on a log intensity scale shows evidence of a Burstein-Moss shift altering the absorption onset as a function of cation composition. Absorption onset varies from 1.0--1.4 eV as carrier density increases by almost two orders of magnitude and both occur while cation composition varies from 0.60--0.40 Zn/(Zn+Sn). Due to the limit of our detector for films of this thickness, any absorption below $\sim$10\hi{3} cm\hi{-1} could not be observed. Panels (a), (c), (d), and (f) show the individual traces at the extremes of panels (b) and (e).}
\end{figure}

\subsubsection{Light Absorption}

Fig. 3 shows the absorption coefficient plotted on a log scale as a function of Zn fraction on the cation site at two different growth temperatures: 160\degree C in panels (a)--(c), and 230\degree C in panels (d)--(f). The data and absorption trends shown in Fig. 3 are representative of the broader range of libraries studied in this work for growth temperatures above $\sim$100\degree C. Absorption coefficient color maps are shown in Figs. 3b and 3e, in which absorption coefficient values above 3~x~10\hi{4} cm\hi{-1} are displayed in yellow and below 8~x~10\hi{3} cm\hi{-1} are displayed in black. The panels that bookend Figs. 3b and 3e are traditional representations of absorption coefficient versus photon energy, also plotted on a log scale. Panels (a) and (c) correspond to the Zn-rich (0.60) and Sn-rich (0.40) limits on the color map in panel (b), respectively. The same description is true of panels (d) and (f). 

As film composition moves from Sn-rich to Zn-rich, free carrier absorption decreases until none is detectable for films grown at 230\degree C with 0.60 Zn/(Zn+Sn) (Fig. 3d). Free carrier absorption is visualized as a rise in absorption coefficient at low photon energy preceding the absorption edge, and is observed most strongly in Figs. 3c and 3f. Decreasing free carrier absorption coincides with two important trends: (1) absorption edge shift from $\sim$1.4 eV to 1.0 eV, and (2) decreasing carrier density. Such correlation suggests a Burstein-Moss shift\cite{wu2004} causing an increase in the apparent bandgap as a result of conduction band filling. Both stoichiometric and Sn-rich films had carrier densities 1--2 orders of magnitude higher than their Zn-rich counterparts, consistent with a Burstein-Moss shift. Considering that degree of cation disorder (based on observation of wurtzite structure by XRD) stayed nominally constant as cation composition varied from 0.40--0.60 Zn/(Zn+Sn), it is unlikely that the effects of an order-disorder transition are convoluted with the Burstein-Moss effect, as observed in prior work.\cite{feldberg2013} The optical absorption edge for films with no detectable free carrier absorption was consistently found to be 1.0 eV; in good agreement with the calculated bandgap for cation-disordered ZnSnN\lo{2}.\cite{feldberg2013, lahourcade2013} 

An experimental effective mass (m*) was calculated from the carrier density and absorption edge data presented in Figs. 2 and 3. Using a zero Kelvin approximation,\cite{kittel} the following equation can be rearranged to obtain a relationship between the slope of the fitted line and m* by taking $E_f$ to be the measured absorption onset.

\begin{equation}
	E_f=\frac{\hbar^{2}}{2m^*}\left(\frac{3\pi^{2}N}{V}\right)^{2/3}
\end{equation}

\begin{figure}[h!]
	\centering
		\includegraphics[height=7.5cm]{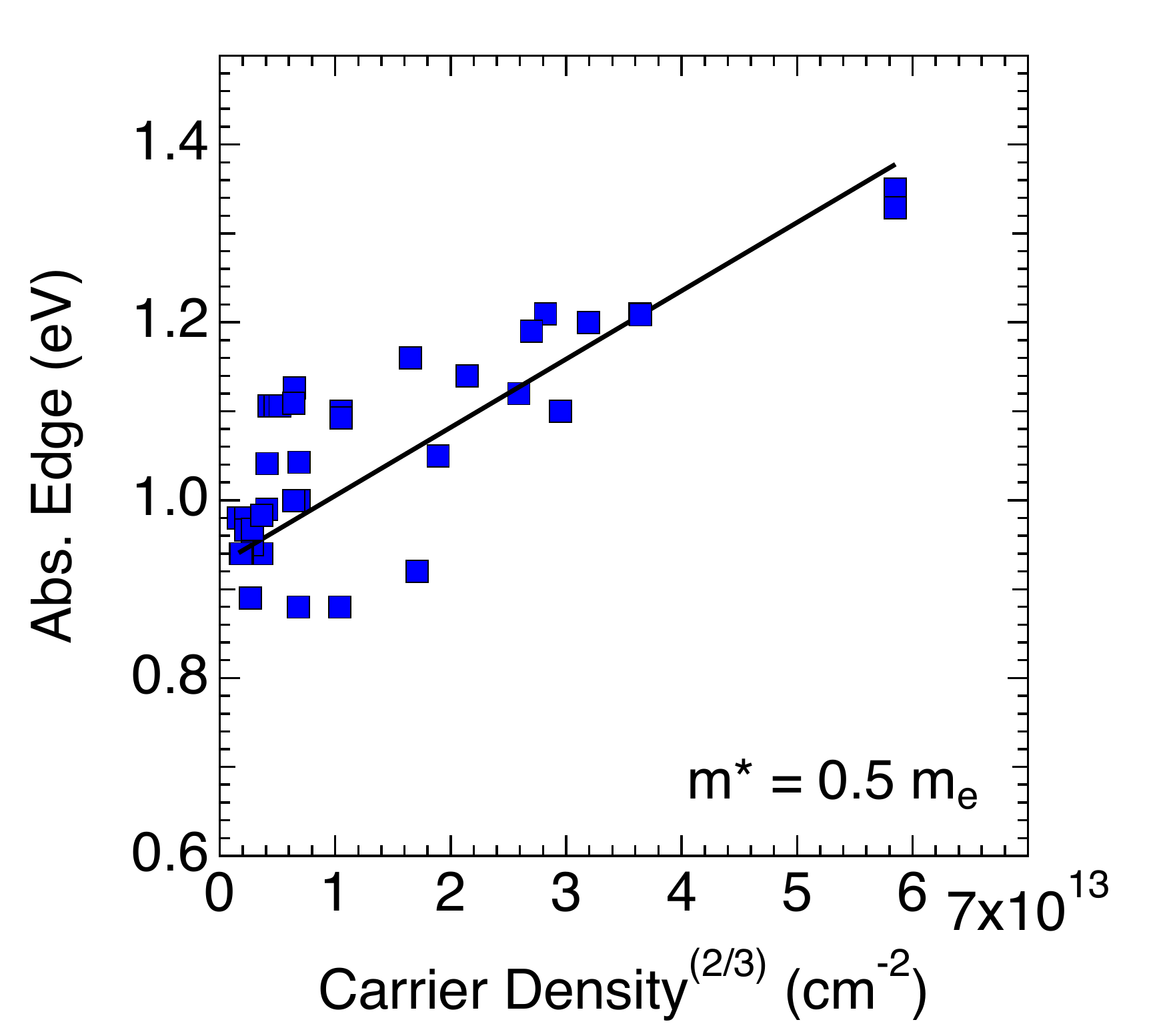}
	\caption{Due to the Burstein-Moss shift observed in this material, the data from Figs. 2 and 3 can be used to calculate an experimental effective mass by plotting experimental absorption edge against measured carrier density raised to the 2/3 power. An experimental effective mass (m*) of 0.5m\lo{e} was extracted from the absorption coefficient and carrier density data presented above. This was done by performing a linear fit of the data and setting the slop of that line equal to the slop of Eq. 1, and solving for m*.}
\end{figure}

Absorption edge was plotted against carrier density raised to the 2/3 power for samples displayed in Fig. 2b, and a linear fit of the resulting plot was performed to determine a slope (Fig. 4). The slope of the fitted line in Fig. 4 was found to be 7.68 x $10^{-15}$ eV cm\hi{2}, and yielded m* = 0.5m\lo{e}. This value is greater than the previously calculated value of 0.1m\lo{e} for ordered, Pna2\lo{1} orthorhombic ZnSnN\lo{2}.\cite{chen2014,lahourcade2013} This discrepancy is unsurprising given that a larger effective mass is expected for a wurtzite crystal structure based on prior calculations.\cite{feldberg2013} The larger effective mass might also be a result of the approximate nature of this analysis. However, the effective mass extracted from Fig. 4 is of the appropriate order of magnitude and is \textless~1m\lo{e}, which is qualitatively consistent with previous calculations.\cite{chen2014,lahourcade2013}

\begin{figure}[ht!]
	\centering
		\centerline{\includegraphics[width=9.5cm]{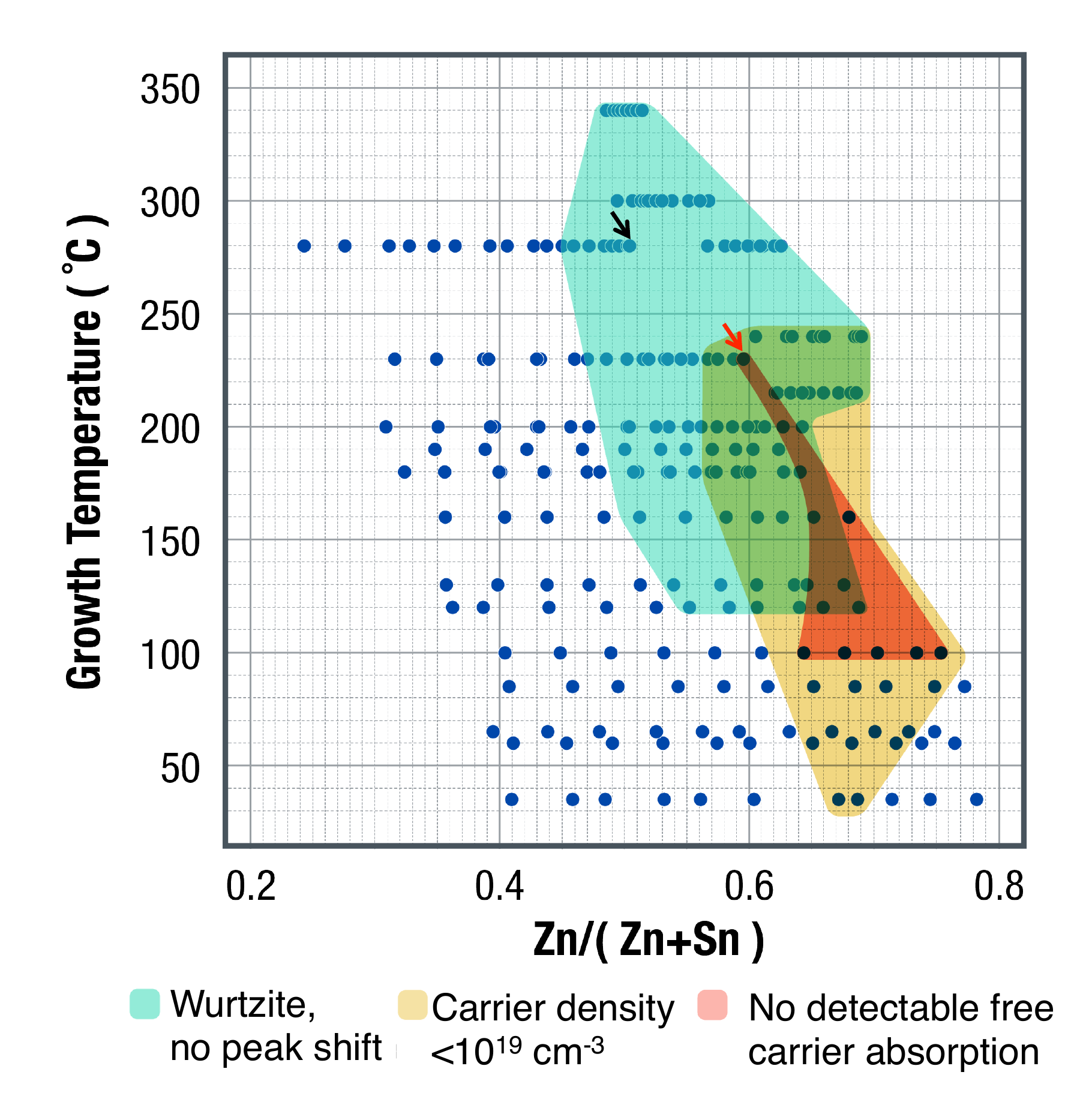}}
	\caption{Summary of data collected via combinatorial experiments. Each point on the figure represents one of 44 data points collected from each of six combinatorial libraries. The teal region indicates deposition conditions for which wurtzite ZnSnN\lo{2} films could be grown without peak shift or broadening by XRD. The yellow region gives the conditions for growing films with \emph{n}-type carrier density below 10\hi{19} cm\hi{-3}. The orange region gives the boundaries for growing films with no free carrier absorption below the absorption edge (\emph{i.e.} carrier density $\leq$ 4 x 10\hi{18} cm\hi{-3}), and is a subset of the yellow region. The red arrow indicates films that exhibited the largest grain size overall, 70 nm diameter on average, and are considered to be the optimal films grown in this work. The black arrow indicates a film in which XRD peaks consistent with orthorhombic crystal structure were observed.}
\end{figure}

\subsubsection{Summary of Combinatorial Work}

The diagram shown in Fig.~5 is a property map summarizing the results of our combinatorial work on ZnSnN\lo{2}. Each dark blue dot displayed in the figure represents a particular set of growth conditions examined in this work. The shaded overlaid shapes indicate boundaries in growth temperature-composition space where particular material properties were observed. The teal shaded region indicates the growth temperature and cation composition range we identified for growing wurtzite ZnSnN\lo{2} films without peak shift or broadening by XRD that would indicate the presence of secondary phases. The overlapping yellow region gives the boundaries for obtaining films with \emph{n}-type carrier density below 10\hi{19} cm\hi{-3}. Finally, the orange triangle gives the boundaries for growing films that exhibit no free carrier absorption below the absorption edge (\emph{i.e.} carrier density $\leq$ 4x10\hi{18} cm\hi{-3}), and should be considered a subset of the yellow region. Samples grown at conditions beyond these boundaries exhibited one or a combination of the following: (1) under-dense morphology or non-columnar grain growth by SEM, (2) amorphous character or possible secondary phases by XRD, and/or (3) \emph{n}-type carrier densities upwards of 10\hi{19} cm\hi{-3}. Particularly, films grown at 400\degree C showed regions of no net deposition or regions of metallic zinc and tin mixed with the binary Zn\lo{3}N\lo{2} and Sn\lo{3}N\lo{4} parent phases. The red arrow at 230\degree C and $\sim$0.60 Zn/(Zn+Sn) indicates films that exhibited the largest grain size overall, 70 nm diameter on average, and are considered to be the optimal films grown in this work. The black arrow indicates films in which XRD peaks consistent with orthorhombic crystal structure were observed, although these films still exhibited carrier density \textgreater 10\hi{20} cm\hi{-3}. Given the considerable difficulty previously experienced in determining the fundamental properties of ZnSnN\lo{2}, the property map displayed in Fig. 5 can provide a platform for accelerated development of this material in the future. 

\subsection{Advanced Characterization}

By using a combinatorial approach, we have been able to put forth guidelines for growth of ZnSnN\lo{2} thin films with promising properties for photovoltaics. The principle advantage of the combinatorial approach is manifest in our ability to rapidly screen a wide range of deposition conditions in the search for regions of desirable material properties. However, combinatorial-based characterization can only go so far in illuminating the finer details of materials grown under conditions that yield interesting properties. To balance this, advanced characterization techniques are necessary to provide a more focused view of the regions of interest identified by our initial combinatorial work. In the following section, we delve deeper into the properties of a small selection of samples in order to better understand (1) the challenges that remain to be addressed in future work, and (2) possible routes toward optimization of ZnSnN\lo{2} moving forward.

\begin{figure}[ht!]
	\centering
		\includegraphics[width=9cm]{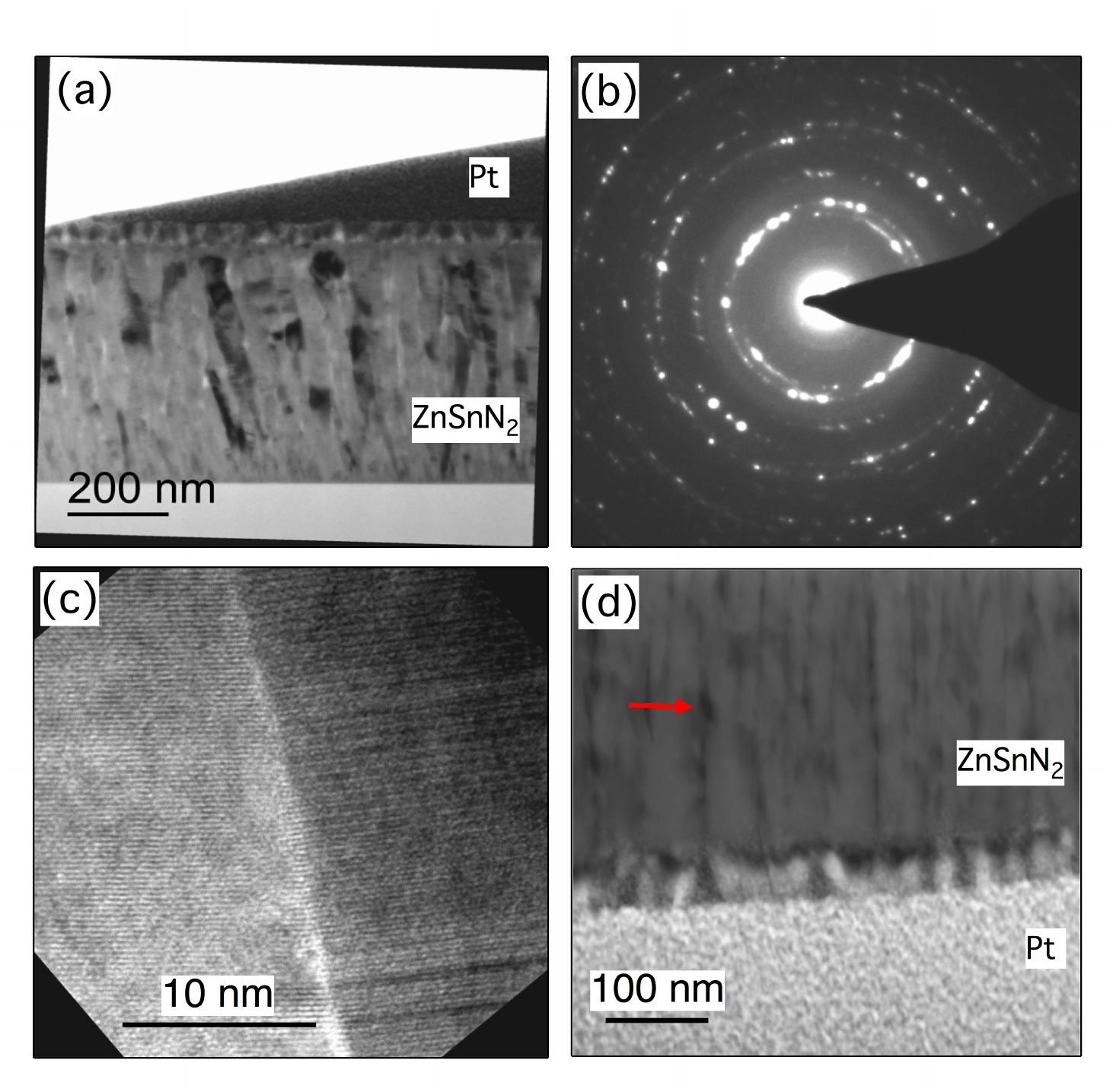}
	\caption{Bright field TEM (a) of a stoichiometric sample grown at 300\degree C shows columnar grain structure, consistent with SEM. A corresponding TED pattern (b) shows polycrystalline morphology and some preference for growth along [0001]. A top-down HRTEM image (c) for the same sample shows atomic-scale crystallinity and grain boundaries free from precipitates. Panel (d) is an HAADF image that when paired with EDX showed evidence of nanovoids at the grain boundaries (indicated by a red arrow).}
\end{figure}

To gain deeper insight into the properties and challenges of our most interesting ZnSnN\lo{2} films, more advanced characterization techniques were applied to selected samples in a non-combinatorial way. We selected samples grown at temperatures in the range of 230--300\degree C that were all either stoichiometric or zinc-rich in composition. These samples represent regions in which the lowest carrier density, sharpest XRD peaks, and/or largest grain size were observed by combinatorial characterization. 

\subsubsection{Crystal Structure}	

Columnar grain structure, observed by SEM, was confirmed by bright-field TEM. A representative image taken from a stoichiometric film grown at 280\degree C is shown in Fig. 6a. Figs. 6b-d, showing transmission electron diffraction (TED), high resolution TEM (HRTEM), and high angle annular dark field (HAADF) images, respectively, were also taken from the same sample. The TED pattern shown in Fig. 6b was used to calculate interplanar spacings corresponding to the six smallest-diameter diffraction rings. These were found to be consistent with those calculated previously\cite{lahourcade2013} for wurtzite ZnSnN\lo{2} with Zn and Sn distributed randomly on the cation sublattice (given in Table S1 of ESI). The pattern in Fig. 6b is consistent with polycrystalline film morphology, which is consistent with XRD analysis, and exhibits evidence of some preferential orientation along the [0001] direction (c-axis).

HRTEM of a representative grain boundary is shown in Fig. 6c. Grain boundaries of stoichiometric and 0.55 Zn/(Zn+Sn) samples were free of precipitates, such as metallic Zn or Sn, and were crystalline at their interfaces. Small regions of high dislocation density were also observed within grain interiors and away from the grain boundaries. These regions were no more than 5 nm across and sometimes less than 1 nm across, and were only found in about half the grains imaged. Here again, these dislocations are more likely artifacts of a fast deposition rate and growth on non-oriented substrates, and may be avoided under more stringent deposition conditions or through post-growth annealing.
	
HAADF imaging together with EDX line profiles revealed the presence of nano-scale voids in the stoichiometric film shown in Fig. 6. These voids are visible in Fig. 6d as small dark regions between columnar grains (indicated by a red arrow). In addition to finding nanovoids, grain size was found to be no larger than 50 nm in diameter for samples imaged by TEM. Indeed, grains with 70 nm diameter or smaller were consistently observed by all imaging techniques used in this study (also see Fig. 1). Nanoscale voids and very small grain size when preparing films by sputtering clearly represent challenges to be addressed as research on ZnSnN\lo{2} moves forward. Previous works using sputtering\cite{lahourcade2013, deng2015} have also reported small grains, while MBE growth\cite{feldberg2013, feldberg2014} on (111) yttria-stabilized zirconia (YSZ) substrates unsurprisingly produced single-crystal ZnSnN\lo{2}. One report even showed preliminary evidence that annealing sputtered ZnSnN\lo{2} films increases grain size, although the maximum size reported was still only 7.5 nm (based on Scherrer analysis) when the initial grain size was 5.8 nm.\cite{deng2015} Taken together, these examples from literature support the conclusion that the challenges mentioned above are related to the sputtering growth method used in the present study and not fundamental to this material.

\begin{figure}[ht!]
	\centering
		\includegraphics[width=5cm]{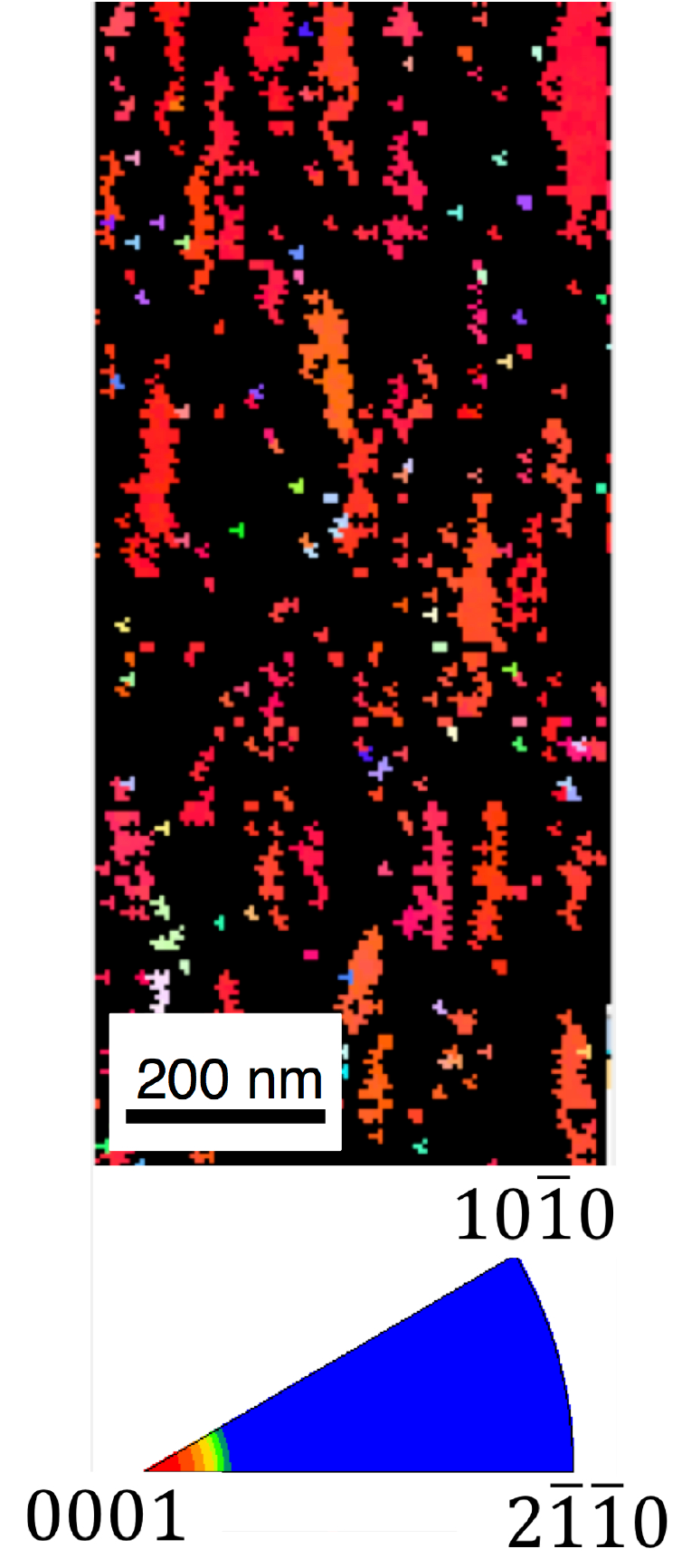}
	\caption{EBSD grain orientation map corresponding to the stoichiometric sample imaged with TEM in Fig. 5. Kikuchi patterns were indexed with high coincidence using a hexagonal lattice. The preferred orientation was found to be along [0001] (shown in red), which is consistent with the TED pattern in Fig. 5. Dark regions indicate regions where no, or low-quality Kikuchi patterns were observed.}
\end{figure}

An electron back scatter diffraction (EBSD) grain orientation map of the stoichiometric ZnSnN\lo{2} film corresponding to the XRD pattern displayed in the top panel of Fig. 1a is shown in Fig. 7. This region exhibited XRD peaks that are consistent with the Pna2\lo{1} orthorhombic structure calculated in Ref. [6].\nocite{lahourcade2013} Kikuchi patterns for this region were indexed using a hexagonal lattice and routinely yielded coincidence indices of greater than 0.4, meaning very good agreement with hexagonal crystal structure. It is important to understand that the orthorhombic structures predicted\cite{paudel2008, punya2011pssc, feldberg2013, lahourcade2013} for ZnSnN\lo{2} in literature are examples of higher-level symmetry superimposed on a fundamentally hexagonal lattice. When the cation sub-lattice in ZnSnN\lo{2} attains a high degree of ordering a larger, orthorhombic unit cell must be drawn to fully capture the translational symmetry of the more ordered crystal structure. A detailed depiction of how the orthorhombic unit cell relates to the underlying hexagonal lattice in ZnSnN\lo{2} is given Ref. [2].\nocite{punya2011pssc}

The most prevalent grain orientation observed in Fig. 7 is (0001), shown in red with \textgreater95\% confidence, and no other orientation was so clearly favored. This is consistent with the TED results of Fig. 6b that also revealed a preference for grains to grow along [0001]. Regions of black in Fig. 7 represent areas with no or low-quality Kikuchi patterns, which could not be reliably indexed by the EBSD software. Although absence of Kikuchi patterns is often indicative of amorphous material, this explanation is not consistent with the HRTEM and TED data collected for these films, in which no amorphous character was observed. However, grain size in these films was always smaller than the spot size of the incident electron beam used for the EBSD measurements, and this may have contributed to the difficulty in obtaining Kikuchi patterns in some places. A combination of the small grain size in this sample, which approached the lateral resolution of EBSD, and observed electron-beam damage of the surface probably contributed to the difficulty in obtaining good EBSD data over the whole analyzed area. Moreover, the presence of nanovoids and small regions of high dislocation density found by HAADF imaging and HRTEM, respectively, also likely contributed to the dark regions observed in Fig. 7.

Finding (0001) preferential growth is interesting since these films were grown on amorphous glass and might be expected to exhibit no orientational preference. However, hexagonal ZnO is known to routinely exhibit (0001) texturing on glass, suggesting growth in the c-axis direction may simply be a feature of materials with a hexagonal lattice.\cite{zakutayev2013} This finding suggests that growth on lattice-matched (0001)-oriented hexagonal substrates is a promising route forward for optimizing the morphology of these films. This finding is also consistent with previous works achieving c-axis oriented ZnSnN\lo{2} thin films on (111) yttria-stabilized zirconia, GaN, or c-plane sapphire substrates.\cite{feldberg2013, lahourcade2013, feldberg2014}

\subsubsection{Transport}

An inverse relationship between carrier density and mobility was observed within the optimal range of deposition parameters we identified by combinatorial studies (Fig. 8). This relationship suggests ionized defects may be a dominant source of charge carrier scattering in this material. Typically, this sort of relationship would be expected of temperature-dependent Hall effect data, but we observe it as a function of increasing off-stoichiometry (Zn-richness). This is an intriguing parallel to draw, because in the context of temperature-dependent Hall, lowering the temperature decreases the energy available to activate defects and simultaneously decreases the concentration of charged scattering centers. In our case (Fig. 8), we appear to be losing charged scattering centers as Zn-richness increases. This suggests the formation of defect complexes as the film composition moves away from stoichiometric.

\begin{figure}[ht!]
	\centering
		\includegraphics[width=9cm]{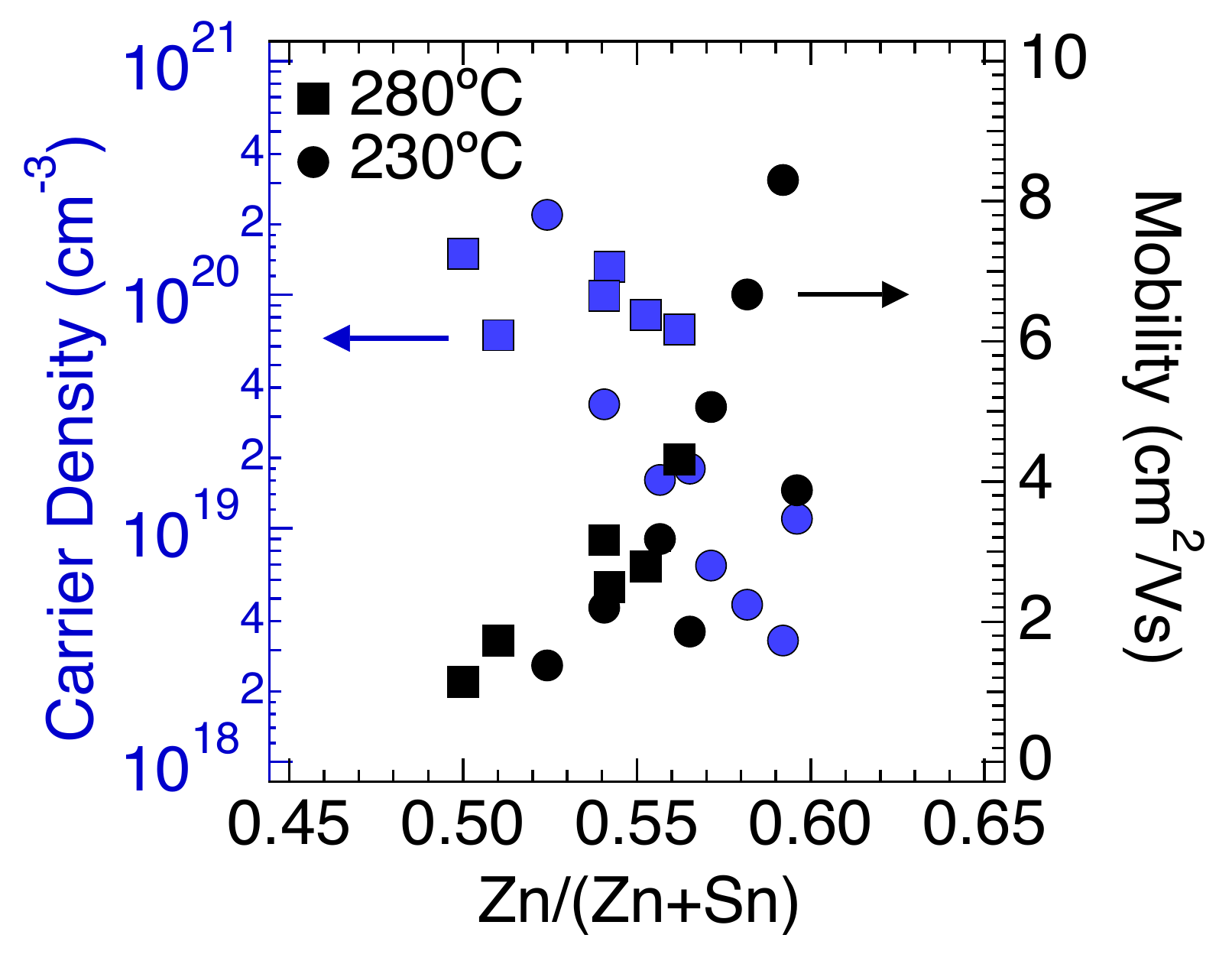}
	\caption{An inverse relationship between carrier density (blue markers) and mobility (black markers) measured by room temperature Hall effect was observed for samples with 0.50--0.60 Zn/(Zn+Sn) and grown between 200--300\degree C. This suggests ionized defect scattering limits mobility for films in this work, and also suggests the formation of defect complexes as films become increasingly Zn-rich.}
\end{figure}

To gain insight into the possible identity of these defect complexes, RBS analysis was performed on a nominally stoichiometric sample grown at 230\degree C on (100) Si. By integrating the area under each region in the RBS spectrum (Fig. 9), we found 4\% oxygen incorporation within the error of the integration procedure. This level of oxygen incorporation would lead to $\sim$5 x 10\hi{20} free electrons/cm\hi{3} for stoichiometric films, which is in good agreement with the measured carrier density in Fig. 2. If we assume all oxygen impurities take the form of O\lo{N} defects that each result in one extra electron, then 4\% oxygen would require 4\% compensation to yield non-degenerate carrier density. The films in Fig. 8 required 10\% excess Zn to reach a compensation with carrier density of 3 x 10\hi{18} cm\hi{-3}, which leaves 6\% compensation unaccounted for. Assuming \emph{n}-type degeneracy comes from the anion sublattice (Zn\lo{i} is high in energy, as described in Section 3.1.2), V\lo{N} defects must also be present. At the most, three extra electrons are donated for each V\lo{N}, which leads to at least 2\% V\lo{N} concentration. Based on this analysis, we propose that defect complexes in off-stoichiometric Zn\lo{1+x}Sn\lo{1-x}N\lo{2} must involve O\lo{N} and V\lo{N} defects. However, more in depth analysis of the defect physics of the full Zn-Sn-N system calls for new theoretical methods that go beyond the simple point-defect model so far applied to cation-ordered ZnSnN\lo{2}.

\begin{figure}[ht!]
	\centering
		\includegraphics[width=9.5cm]{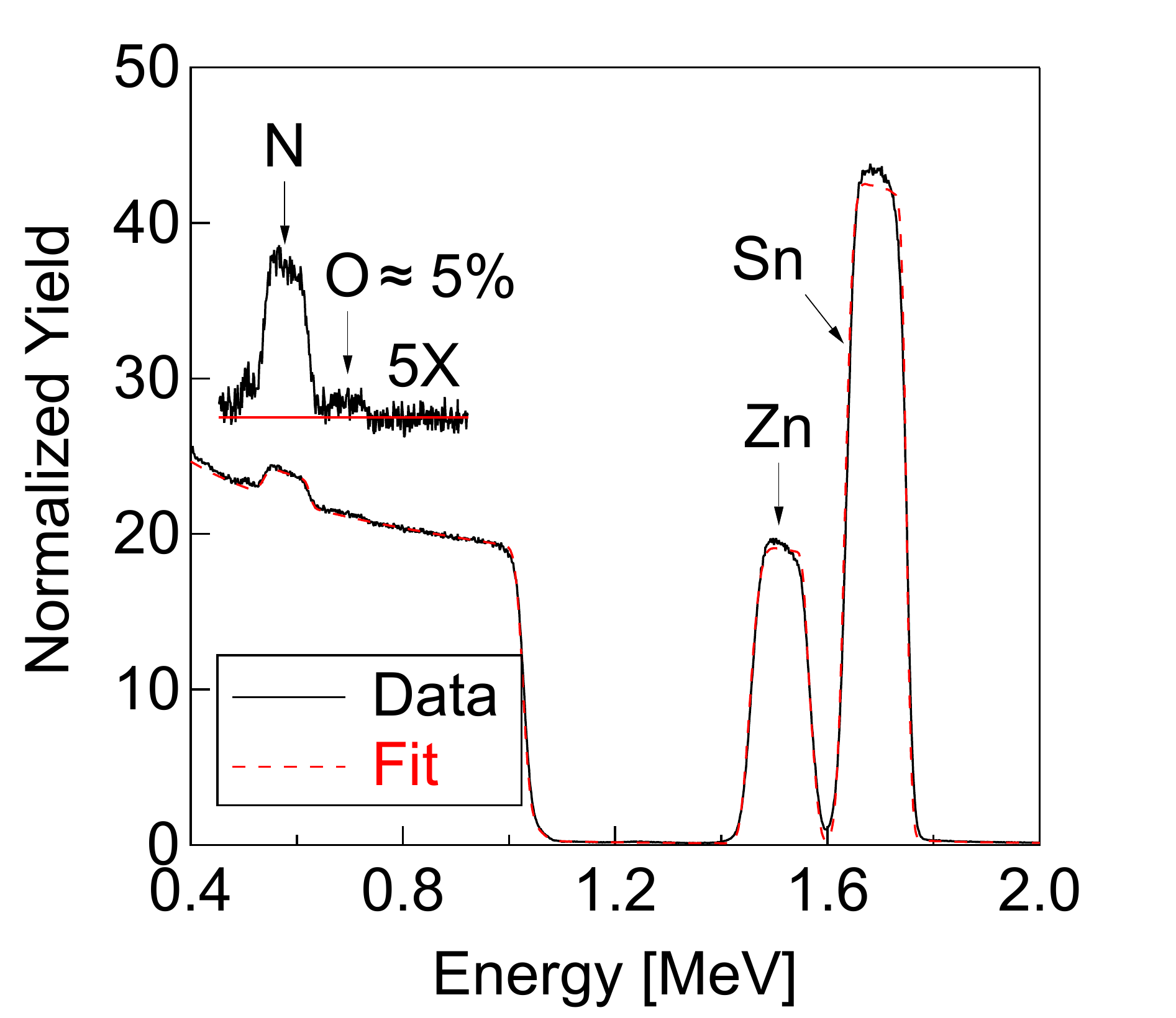}
	\caption{Oxygen impurity concentration in a nominally stoichiometric film grown at 230\degree C was determined by RBS to be 4\%. Assuming every oxygen forms an O\lo{N} defect, this level of oxygen impurities suggests at least 2\% V\lo{N} defects to satisfy the 10\% Zn-rich composition for which \emph{n}-type carrier density of 3 x 10\hi{18} cm\hi{-3} was observed.}
\end{figure}

The role of defect interactions has become increasingly appreciated in multinary tetrahedrally-bonded semiconductors, and suggests the isolated point defects discussed above may be an incomplete picture. To continue the discussion from a defect interaction perspective, we now consider the ``motif clustering'' (\emph{e.g.} compositional inhomogeneity) model recently proposed by Ref. [33].\nocite{zawadzki2015} According to the framework outlined therein, ZnSnN\lo{2} should only have one motif in its ground state: N-Zn\lo{2}Sn\lo{2}, meaning N coordinated by two Zn and two Sn atoms. This motif satisfies the octet rule for the anion, considering that each cation shares its charge among four neighboring nitrogen atoms. There are five possible motifs overall: N-Zn\lo{2}Sn\lo{2}, N-Zn\lo{3}Sn, N-ZnSn\lo{3}, N-Zn\lo{4}, and N-Sn\lo{4} (listed in order of least deviation to most deviation from the octet rule). If we apply this paradigm to off-stoichiometric Zn\lo{1+x}Sn\lo{1-x}N\lo{2} films, we find a possible explanation for our ability to grow wurtzite ZnSnN\lo{2} at compositions ranging from 0.45--0.70 Zn/(Zn+Sn). Under conditions of high Zn chemical potential (larger Zn flux than Sn flux), the motif clustering model suggests that the Zn-rich motif N-Zn\lo{3}Sn will form in addition to the ground state N-Zn\lo{2}Sn\lo{2} motif. The same can be said of growth under Sn-rich conditions (0.45 Zn), but with the formation of N-ZnSn\lo{3} motifs. Formation of higher-energy motifs has been shown to lead to motif clustering in other multinary, tetrahedrally bonded semiconductors.\cite{baranowski2015, zawadzki2015} In those works, motif clustering was implicated in causing charge localization, leading to ionized defect scattering. Considering that the measured average mobility of the films in this work is lower than would be expected from the light effective mass calculated previously,\cite{lahourcade2013} the motif clustering model may provide insight into the possible origin of this discrepancy.

\subsubsection{Photoluminescence}

At Zn-rich compositions, films consistently exhibited a broad peak in low temperature PL intensity in the range 1.35--1.5 eV (Fig. 10). The spectra in Fig. 10 were taken from the Zn-rich region of a 300\degree C sample, and are representative of other Zn-rich films grown at lower temperatures. The energy range in which the PL peaks occur coincides with the energy at which another broad PL peak was observed in a previous work\cite{quayle2013} at room temperature. This previous study attributed their PL peak at 1.4 eV to defect luminescence. However, the average peak energy in Fig. 10 is $\sim$0.4 eV above the corresponding room temperature absorption onset of 1.0 eV observed for similar films (see Fig. 3). Defect luminescence is usually observed at energies below the optical bandgap in PL measurements, so observing luminescence above the optical gap suggests that this signal might not be consistent with a defect state. It is possible that the bandgap of cation-disordered ZnSnN\lo{2} has a large temperature-dependence, but systematic investigation into this possibility will be the topic of a future study. Given that desirable properties such as low carrier density and sharp absorption edge were observed for Zn-rich films, confirmation that PL signal could be obtained from such samples is promising.

\begin{figure}[h!]
	\centering
		\includegraphics[width=8cm]{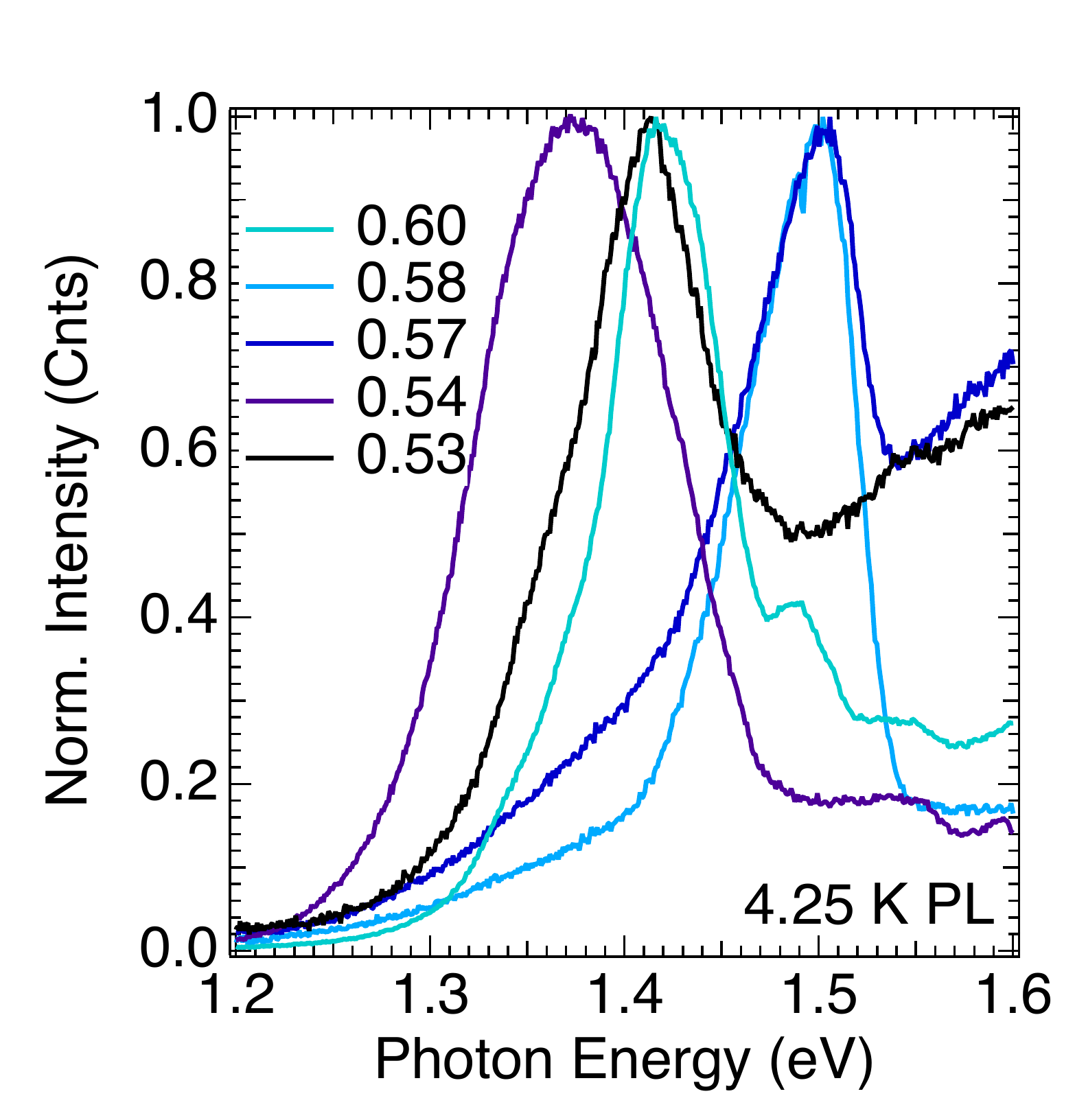}
	\caption{Normalized PL intensity plotted as a function of incident photon energy. PL signal has been observed at 4.25 K for nominally Zn-rich films in which many of the optimal properties reported in this work have been found. Consistently, PL intensity was observed in the range 1.35--1.5 eV.}
\end{figure}

\section{Summary and Conclusions}

In this work, we have presented a comprehensive investigation into the Zn-Sn-N material system. Using a combinatorial approach paired with advanced characterization we have been able to address several of the primary materials challenges that previously hindered further development of ZnSnN\lo{2} for optoelectronics. These obstacles include degenerate carrier density, uncertainty with respect to the fundamental gap, and control of the cation order parameter. By varying the degree of cation off-stoichiometry, we achieved an \emph{n}-type carrier density of 1.8~x~10\hi{18} cm\hi{-3}, which is the lowest carrier density yet reported for as-deposited films. Carrier density was found to be inversely proportional to mobility as cation composition varied from 0.50--0.60 Zn/(Zn+Sn), indicating that defect complexes have a significant effect on the carrier concentration of ZnSnN\lo{2}. Based on RBS analysis, which showed 4\% oxygen incorporation in stoichiometric films, we propose these defect complexes likely involve O\lo{N} and V\lo{N} defects. Control of the carrier density additionally provided insight into the fundamental gap and the conduction band effective mass of cation-disordered ZnSnN\lo{2}. Analysis of the Burstein-Moss shift for cation-disordered samples with carrier densities ranging from 10\hi{18}--10\hi{20} cm\hi{-3} revealed a fundamental band edge of 1.0 eV and a conduction band effective mass of 0.5m\lo{e}. In addition, the absorption coefficient for non-degenerate, cation-disordered films was found to rise strongly from 1.0 eV, suggesting disordered ZnSnN\lo{2} is a candidate absorber for thin film photovoltaics. Finally, we observed evidence of controlling cation order parameter through substrate temperature during growth, with higher growth temperature enabling the transition to the ordered orthorhombic (Pna2\lo{1}) structure. Overall, we conclude that the results presented herein not only provide insight into the fundamental properties of the Zn-Sn-N material system and highlight the potential to utilize ZnSnN\lo{2} as a PV absorber, but also help pave the way for more rapid advancement of research into ZnSnN\lo{2} in the future.

\section{Acknowledgements}

This work was supported by the U.S. Department of Energy as a part of the Non-Proprietary Partnering Program under Contract No. De-AC36-08-GO28308 with the National Renewable Energy Laboratory. A.N.F. was supported by the Renewable Energy Materials Research Science and Engineering Center under Contract No. DMR-0820518 at the Colorado School of Mines. Thanks to Bobby To, Patricia C. Dippo, and Adam Stokes at the National Renewable Energy Laboratory (NREL) for SEM, PL, and TEM sample prep. Thanks to Joshua Bauer at NREL for the property diagram illustration. 

\balance
\bibliographystyle{rsc} 
\bibliography{ZnSnN2} 

\end{document}